\documentclass[natbib]{svproc}

\usepackage{url}
\usepackage{natbib}
\usepackage{graphicx}
\usepackage{amsfonts}
\usepackage{amsmath}
\usepackage{wrapfig}
\usepackage{subcaption}
\usepackage{hyperref}
\usepackage{booktabs}
\usepackage{float}
\floatstyle{plaintop}
\restylefloat{table}

\newcommand{\change}[1]{#1}

\begin{document}
\mainmatter              %
\title{Generalising Multi-Agent Cooperation through Task-Agnostic Communication}
\titlerunning{Task-Agnostic Communication}  %
\author{Dulhan Jayalath*\inst{1}, Steven Morad\inst{2}, \and
Amanda Prorok\inst{2}}
\authorrunning{Dulhan Jayalath et al.} %
\tocauthor{Dulhan Jayalath, Steven Morad, and Amanda Prorok}
\institute{University of Oxford\\
\email{dulhan@robots.ox.ac.uk},\\
\and
University of Cambridge\\
\email{\{sm2558, asp45\}@cl.cam.ac.uk}\\
}
\maketitle              %

\begin{abstract}
Existing communication methods for \textit{multi-agent reinforcement learning (MARL)} in cooperative multi-robot problems are almost exclusively task-specific, training new communication strategies for each unique task. We address this inefficiency by introducing a communication strategy applicable to any task within a given environment. We pre-train the communication strategy \textit{without} task-specific reward guidance in a self-supervised manner using a set autoencoder. Our objective is to learn a fixed-size latent Markov state from a variable number of agent observations. Under mild assumptions, we prove that policies using our latent representations are guaranteed to converge, and upper bound the value error introduced by our Markov state approximation. Our method enables seamless adaptation to novel tasks without fine-tuning the communication strategy, gracefully supports scaling to more agents than present during training, and detects out-of-distribution events in an environment. Empirical results on diverse MARL scenarios validate the effectiveness of our approach, surpassing task-specific communication strategies in unseen tasks. Our implementation of this work is available at \url{https://github.com/proroklab/task-agnostic-comms}.\let\thefootnote\relax\footnotetext{*Work done while author was at the University of Cambridge.}
\end{abstract}

\section{Introduction}

Reinforcement learning has succeeded in a number of robotics applications where classical methods have struggled \citep{ibarz2021, han2023, orr2023}. For multi-robot systems, MARL often promises better scaling to large numbers of agents \citep{christianos2021}, and results in interesting properties like emergent tool use \citep{baker2020}. However, MARL is sample-inefficient, making it costly to deploy to real-world robotics tasks \citep[Section 3.2]{marinescu_prediction-based_2017, zhang_multi-agent_2019}. %

MARL is \emph{partially observable} in the sense that individual agent observations are often insufficient to learn an optimal policy. Rather, we must reason over \emph{all} agent-level observations to find an optimal policy. This partial observability further worsens the limited sample efficiency suffered by single-agent RL \citep{buckman_sample-efficient_2018, yu_towards_2018}. To alleviate these issues in collaborative settings, many approaches utilise communication to share information between agents \citep{foerster_learning_2016, sukhbaatar_learning_2016, das_tarmac_2019, bettini_heterogeneous_2023}. These methods typically use a differentiable strategy, optimising messages with respect to the reinforcement learning objective. 

Thus far, differentiable communication for cooperative multi-agent learning has been entirely task-driven. Previous works have learned communication strategies for solving riddles \citep{foerster_learning_2016}, traffic control \citep{sukhbaatar_learning_2016, das_tarmac_2019}, navigation \citep{das_tarmac_2019, li_graph_2020}, and tasks requiring heterogeneous behaviour \citep{bettini_heterogeneous_2023}. In every example, agents learn a communication strategy, specific to each task that they are expected to solve. Learning such \textit{task-specific} communication strategies is wasteful and inefficient---particularly given the poor sample efficiency of MARL.

\begin{figure}[t]
    \centering
    \includegraphics[trim={0 1.1cm 0.9cm 1.2cm},clip,width=\linewidth]{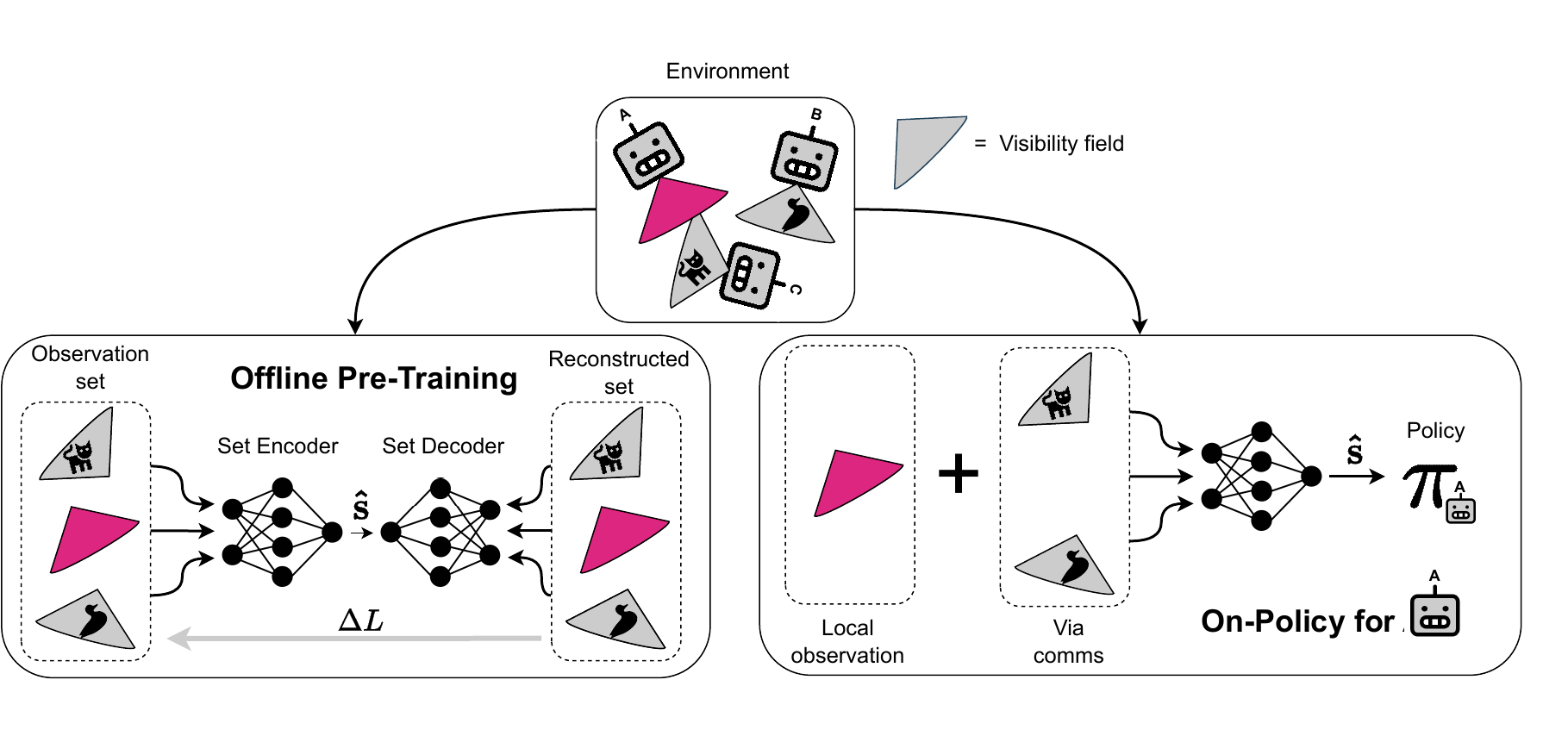}
    \caption{\textbf{Learning and applying a task-agnostic communication strategy in MARL.} \textit{Offline pre-training.} We pre-train a set autoencoder with sets of observations collected from exploring an environment. Since there is no reward signal involved in sampling these observations, the autoencoder learns a task-agnostic representation. When a variable number of agent observations are encoded by the set encoder, the output is a fixed-size latent vector $\mathbf{\hat{s}}$ approximating the Markovian state $s$ in a Dec-MDP. \textit{Policy training.} For each agent, we deploy the pretrained set encoder to encode the global observation (assembled via communication) into $\mathbf{\hat{s}}$ on the fly. We condition the behaviour policy on $\mathbf{\hat{s}}$.}
    \label{fig:fig1}
\end{figure}

\textbf{Approach.} We propose improving the efficiency of communication strategies by learning a \textit{task-agnostic} and \textit{environment-specific} communication strategy (Figure \ref{fig:fig1}).  We differentiate between an environment (a world and its transition dynamics) and a task (an optimisation objective within an environment). In task-specific strategies, even if the environment does not change, agents must learn a new communication strategy for each new task. In contrast, a task-agnostic strategy can be shared across all tasks \textit{within this environment}. This is critical towards designing general-purpose robots as they commonly engage in a variety of tasks in shared environments (e.g. warehouses, homes, cities, etc.).

Our method also brings several auxiliary advantages. Firstly, by utilising a specialised set autoencoder, we enable decoding a \textit{fixed-size latent state} into a \textit{variable-sized set}. This permits training the communication strategy to elegantly support variable numbers of agents and to even scale \textit{out-of-distribution} to more agents than seen during training. Additionally, by comparing pre-training losses to the losses at runtime, it is possible to detect out-of-distribution disturbances in an environment (e.g. adversarial agents and unsafe environment states). Finally, our approach is \textit{grounded} in the environment, resulting in messages which have specific meaning for any task within this environment.

\textbf{Contributions.} We develop a method for learning general, task-agnostic communication strategies in multi-robot teams that supports variable numbers of agents. We provide two proofs which demonstrate that (\textit{i}) under mild assumptions, our method guarantees return convergence and (\textit{ii}) when these assumptions are not met, there is an upper bound on the regret. We test our method with experiments on tasks in VMAS \citep{bettini_vmas_2022} and the Melting Pot suite \citep{agapiou_melting_2023}. Our task-agnostic communication strategy outperforms repurposed task-specific communication strategies (i.e. trained with policy losses on one task and deployed on another task in the same environment). Moreover, we provide evidence that performance does not degrade significantly as we scale the number of agents in the system. Lastly, we showcase how our pre-training method can be used to detect out-of-distribution events in the environment.

\section{Preliminaries}

We are interested in a subset of cooperative multi-agent problems known as \textit{decentralised Markov decision processes (Dec-MDPs)}. In the Dec-MDP framework, the environment provides a joint observation $o = \{ o_1,\dots,o_n \}$ from which each agent observes its local observation and decides on an action. Upon taking an action, the global reward function $R$ provides a shared reward for each time step, and each agent receives its next local observation \citep{bernstein_complexity_2002}. A key feature of Dec-MDPs is that the underlying Markov state must be \textit{uniquely defined} by the \textit{joint set of observations} of all agents (i.e. the global observation) \citep{oliehoek_concise_2016}.

Formally, denoting $\Delta(X)$ as the set of probability distributions over the set $X$, a Dec-MDP is defined by a tuple $(n, S, A, T, \mathbb{O}, O, R, \gamma)$ where $n$ is the number of agents, $S$ is the set of states (with initial state $s_0$), $A$ is the set of actions for each agent,  $T : S \times A^n \rightarrow \Delta(S)$ is the state transition probability function $T(s'\mid s,\mathbf{a})$, $\mathbb{O}$ is the set of joint observations, $O : S \times A^n \rightarrow \Delta(\mathbb{O})$ is the observation probability function $O(\mathbf{o} \mid s, \mathbf{a}$), $R : S \times A^n \rightarrow \Delta(\mathbb{R})$ is the global reward function $R(s, \mathbf{a})$, $\gamma$ is the discount factor, and the multi-agent Markov state $s$ is unambiguously determined by $\mathbb{O}$.\footnote{This notation is inspired by \citet{oliehoek_concise_2016} and \citet{ellis_smacv2_2022}.}

\section{Learning Task-Agnostic Communication}

In a Dec-MDP, agent cooperation depends on the Markovian state of the multi-agent team within an environment. This state is \textit{independent} of task specifics, consisting of the current state of the \textit{environment}. Therefore, we define a communication model which reconstructs this environment-specific information, and show how it can be trained without reward guidance to be completely task-agnostic.

\subsection{A Communication Model for Task-Agnostic Cooperation}
\label{sec:commsmodel}

 As the global observation defines the multi-agent Markov state in a Dec-MDP, we define our communication model as one in which agents reconstruct the multi-agent state by reconstructing the joint set of all agents' observations.

Consider a Dec-MDP defined by the tuple $(n, S, A, T, \mathbb{O}, O, R, \gamma)$ with agents $i \in \mathcal{A}_n = \{1, \dots, n\}$. Agents have a communication range $\epsilon$ where if the distance $d(i, j)$ between agents $i$ and $j$ is greater than $\epsilon$ then they cannot share information. Thus, we define the neighbourhood of agent $i$ as $\mathcal{N}_i = \{j \in \mathcal{A}_n \mid d(i, j) \leq \epsilon, j \neq i\} $. In each time step $t$, an agent $i$ receives a set which contains the observations of all agents within $i$'s range,
\begin{equation}
\mathbb{O}_t^{\mathcal{N}_i} = \{ o_t^j \mid \forall j \in \mathcal{N}_i \}.  %
\end{equation}
Let $\mathbb{O}_t$ denote the joint set of all agent observations in time step $t$. With $\mathbb{O}_t^{\mathcal{N}_i}$ and its local observation $o_t^i$, the agent can recover the set of observations of all agents within the communication range of $i$ (including itself) in this time step,
\begin{equation}
\mathbb{O}^i_t = \mathbb{O}_t^{\mathcal{N}_i} \cup \{o_t^i\}. %
\end{equation}

Using an autoencoder, the set $\mathbb{O}^i_t$ is encoded into a task-agnostic latent state $\mathbf{\hat{s}}^i_t$. This latent state is permutation-invariant and is a constrained approximation of the global observation $\mathbb{O}_t = \{o_t^i, \dots, o_t^n\}$ (and therefore, the Markov state) constructed using the information available in $\mathbb{O}^i_t$ and the knowledge of the autoencoder. The advantage of this state over a concatenation of all observations is that it is fixed in size, supporting variable numbers of agents, makes use of the sample efficiency afforded by a permutation-invariant state, and is an efficient compressed representation.

To use this approximation of the multi-agent state in decision-making, we condition the policy of agent $i$ on this latent state. Let $\pi_{\theta_t}$ denote a policy parameterised by weights $\theta_t$. The probability that agent $i$ takes action $a_t^i$ is given by $\pi_{\theta_t}(a_t^i \mid \mathbf{\hat{s}}^i_t, o_t^i)$. The policy is conditioned on the agent's local observation, even though $o_t^i$ is encoded within $\mathbf{\hat{s}}^i_t$, because $\mathbf{\hat{s}}^i_t$ is permutation-invariant. Without $o_t^i$, the policy cannot determine which agent it is reasoning about.

When the latent state $\mathbf{\hat{s}}_t$ perfectly captures the global observation, the policy is guaranteed to converge to a local optimum in the return:

\begin{theorem}
\label{theorem:a}
    A policy gradient method, which conforms to the assumptions in \citep[Theorem 3]{sutton_policy_1999}, conditioned on $\mathbf{\hat{s}}_t$ in a Dec-MDP is guaranteed to converge to a local optimum in the return assuming $\mathbf{\hat{s}}_t$ captures $\mathbb{O}_t$ with zero reconstruction error.
\end{theorem}

\begin{proof}
Consider a Dec-MDP defined by the tuple $(n, S, A, T, \mathbb{O}, O, R, \gamma)$. Define a policy $\pi$ parameterised by $\theta$ that maps the global observation $\mathbb{O}$ to a distribution over joint actions $A$. Formally, $\pi_{\theta}(\textbf{a}_t \mid \mathbb{O}_t)$ represents the probability of taking joint action $\textbf{a}_t$ given global observation $\mathbb{O}_t$ and policy parameters $\theta$. The objective is to optimise the policy $\pi$ to maximise the expected return over a trajectory $\tau$ = ($s_1, \textbf{a}_1, s_2, \textbf{a}_2, \dots, s_T, \textbf{a}_T$), where $s_t$ is the multi-agent Markov state at time $t$. Assume a policy gradient method, such as REINFORCE \citep{williams_class_1987, williams_simple_1992}, to update the policy parameters $\theta$. This requires estimating the gradient of the expected return with respect to $\theta$ in order to update these parameters.

Note that (\textit{i}) policy gradient methods converge to a locally optimal policy in Markov decision processes \citep[Theorem 3]{sutton_policy_1999}, (\textit{ii}) by definition: the joint state $s_t$ in a Dec-MDP is the multi-agent Markov state \citep{bernstein_complexity_2002, oliehoek_concise_2016}, and (\textit{iii}) by definition: this state is jointly fully observable in Dec-MDPs \citep{bernstein_complexity_2002, oliehoek_concise_2016}.

Then, since (\textit{iii}) implies the global observation $\mathbb{O}_t$ uniquely defines $s_t$ \citep{oliehoek_concise_2016}, and by (\textit{ii}) $\mathbb{O}_t$ defines the multi-agent Markov state, since we use a policy gradient method, by (\textit{i}) it is guaranteed to converge to a local optimum as our policy is conditioned on $\mathbb{O}_t$, which is equivalent to the underlying Markov state. When the latent state $\mathbf{\hat{s}}_t$ captures $\mathbb{O}_t$ with zero reconstruction error, this result extends to when the policy is conditioned on $\mathbf{\hat{s}}_t$ instead.
\end{proof}

However, in practice we cannot assume that $\mathbf{\hat{s}}_t$ captures $\mathbb{O}_t$ with no error. To quantify the effect of any error on the return, we can place a bound on the regret: the difference in the expected return achieved if the approximation of the underlying Markov state was perfect.

\begin{theorem}
\label{theorem:b}
    Suppose the policy in a Dec-MDP and its associated value function are Lipschitz continuous. Then the regret of a policy learned from an approximation $\mathbf{\hat{s}}_t$ of the underlying Markov state $\mathbf{s}_t$ is bounded above and this bound is directly proportional to the reconstruction error. \footnote{For this theorem, we treat $s_t$ as a vector  $\mathbf{s}_t$ to decompose its approximation into the true state and error.}
\end{theorem}

\begin{proof}
Consider an identical setting to that stated in the proof of Theorem \ref{theorem:a}. Additionally, define a value function $V_{\pi_{\theta}}$ derived from the policy $\pi_{\theta}$ and let $\epsilon$ be some error in reconstructing the underlying multi-agent Markov state $\mathbf{s}_t$. Thus, $\mathbf{\hat{s}}_t$ can be decomposed into $\mathbf{s}_t + \epsilon$.

Assume that $V_{\pi_{\theta}}$ is $K$ Lipschitz continuous where $K \in \mathbb{R}$. Since $V_{\pi_{\theta}}$ is derived from $\pi_{\theta}$, let us also assume that $\pi_{\theta}$ is Lipschitz continuous.

Then,
\begin{align}
    |V_{\pi_{\theta}}(\mathbf{s}_t) - V_{\pi_{\theta}}(\mathbf{\hat{s}}_t)| &\leq K |\mathbf{s}_t - (\mathbf{\hat{s}}_t)| \\
    |V_{\pi_{\theta}}(\mathbf{s}_t) - V_{\pi_{\theta}}(\mathbf{s}_t + \epsilon)| &\leq K |\mathbf{s}_t - (\mathbf{s}_t + \epsilon)| \\
    &= K|-\epsilon| \\
    &= K|\epsilon|.
\end{align}

Thus, the difference in expected return (regret) between a policy which assumes the underlying Markov state is the true state $s_t$ and one which assumes the underlying Markov state is an approximation $\mathbf{\hat{s}}_t$ due to reconstruction error $\epsilon$ is bounded above by $K|\epsilon|$. Since $K$ is a constant, this bound is directly proportional to the root mean squared error $|\epsilon|$.
\end{proof}

Hence, in a Dec-MDP, the reconstruction error $\epsilon$ is precisely the autoencoder's error in reconstructing $\mathbb{O}_t$ from the latent state $\mathbf{\hat{s}}_t$ as the Markovian state $\mathbf{s}_t$ is uniquely defined by the global observation. In general, Theorem \ref{theorem:b} can be extended to Dec-POMDPs as a bound on the error in estimating the global observation rather than the underlying Markov state.

\subsection{Training a Task-Agnostic Communication Strategy}
\label{sec:trainta}

We posit that we can learn a task-agnostic communication method by pre-training an autoencoder with global observations $\mathbb{O}_t$ from exploration of an environment. If the exploration policy is independent of the reward function, the strategy is task-agnostic. We use either a uniform random policy or uniform random sampling from the observation space. Neither requires knowing a reward function and hence any method learned in this way is task-agnostic.

\begin{wrapfigure}[23]{l}{0.5\linewidth}
    \centering
    \includegraphics[trim={2cm 0.3cm 1.27cm 0.28cm},clip,width=\linewidth]{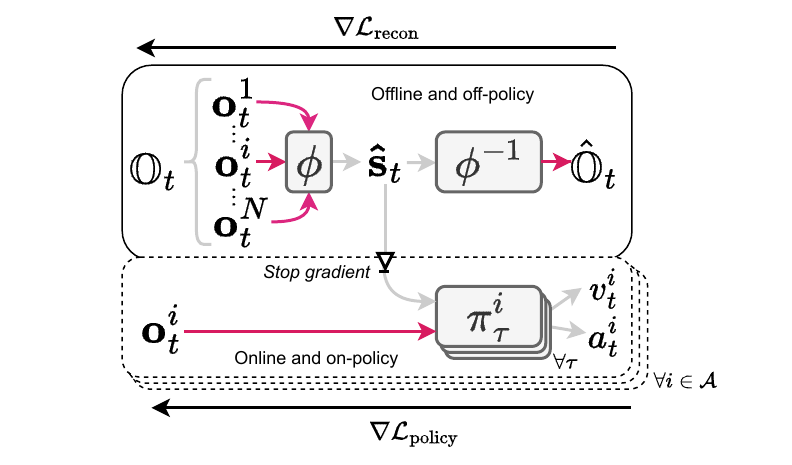}
    \caption{\textbf{Method details.} 
    We collect global observations by exploring the environment in a task-agnostic manner. Using these observations, we pretrain a set autoencoder (with encoder $\phi$ and decoder $\phi^{-1}$) using a self-supervised reconstruction loss. Keeping the autoencoder weights frozen, we train policies $\pi^i$ on various tasks $\tau$. The input to $\pi$ is the approximation of the Markov state $\mathbf{\hat{s}}_t$ and the relevant agent's observation $o^i_t$.}
    \label{fig:method}
\end{wrapfigure}

Figure \ref{fig:method} provides a detailed overview of our method. Given a permutation-invariant set autoencoder with encoder $\phi$ and decoder $\phi^{-1}$, we train the autoencoder with a self-supervised loss
\begin{equation}
    \frac{1}{n} \sum_{t=1}^n l \Bigl( \phi^{-1}(\phi(\mathbb{O}_t)), \mathbb{O}_t \Bigr),
\end{equation}

where $l$ is a function defining a set reconstruction error specified by our choice of autoencoder.

Typically, \textit{graph autoencoders (GAEs)} \citep{tian_learning_2014, wang_structural_2016, kipf_variational_2016} would be ideal to encode sets with permutation invariance. However, we instead use the \textit{permutation-invariant set autoencoder (PISA)} \citep{kortvelesy_permutation-invariant_2023} because, unlike many GAEs, this architecture allows decoding a variable-sized set using a fixed-size latent state. In other words, no matter the number of agents or the corresponding cardinality of the set $\mathbb{O}_t$, the dimension of the latent state $\mathbf{\hat{s}}_t$ is constant. This property is highly desirable as it allows a trained encoder to scale as agents are added or removed from the environment. If pre-trained on global observations, it also enables the autoencoder to approximate the global observation even when some observations in the multi-agent team are missing. %

While many environments emit two-dimensional pixel observations, PISA encodes feature vectors into permutation-invariant states. Given this dichotomy, when required, we also pre-train a convolutional autoencoder on each element of $\mathbb{O}_t$ to encode each pixel observation $o_t^i$ into a feature vector $\mathbf{v}_t^i$. Thus, when an image encoder is necessary, a set of these feature vectors $\mathbb{V}_t$ is the input to our set autoencoder rather than $\mathbb{O}_t$ directly.

\setlength\intextsep{1pt}

\section{Experiments \& Discussion}

We propose three experiments. The first shows that a task-agnostic communication strategy is more effective than a task-specific strategy when presented with a novel task. It also verifies that our proposed strategy outperforms a baseline that does not use communication. Our second experiment validates the claim that our method elegantly handles variable numbers of agents. It shows how our method fares as more agents are introduced, going out-of-distribution with respect to the number of agents seen during pre-training. The final experiment demonstrates that, by comparing pre-training autoencoder losses to the losses during policy training, we can detect out-of-distribution events in the environment.

\subsection{Experimental Setup}
\label{sec:setup}

Our experiments focus on two MARL suites. Firstly, Melting Pot \citep{agapiou_melting_2023} is a suite of 2D, grid-based, discrete multi-agent learning environments, providing scenarios that can test a variety of types of coordination focusing on \textit{social dilemmas}. In this type of scenario, problems are a mixture of competition and cooperation so many tasks are not fully cooperative and accordingly, cannot be Dec-MDPs. To address this, we sum all individual agent rewards emitted by the base Melting Pot task into a shared global reward. This ensures each task is fully cooperative.

To supplement Melting Pot, we also study taasks in the \textit{vectorised multi-agent simulator (VMAS)} \citep{bettini_vmas_2022}: a 2D, continuous-action framework designed for benchmarking MARL. Together, these two suites provide a comprehensive study, as they cover both visual and vector observations, discrete and continuous action spaces, sparse and dense rewards, and different forms of cooperation requirements, ranging from high-level to low-level collaboration strategies.

\setlength\intextsep{0pt}
\begin{wrapfigure}[26]{r}{3.25cm}
    \includegraphics[trim={2.05cm 0 0 0},clip,width=3.25cm]{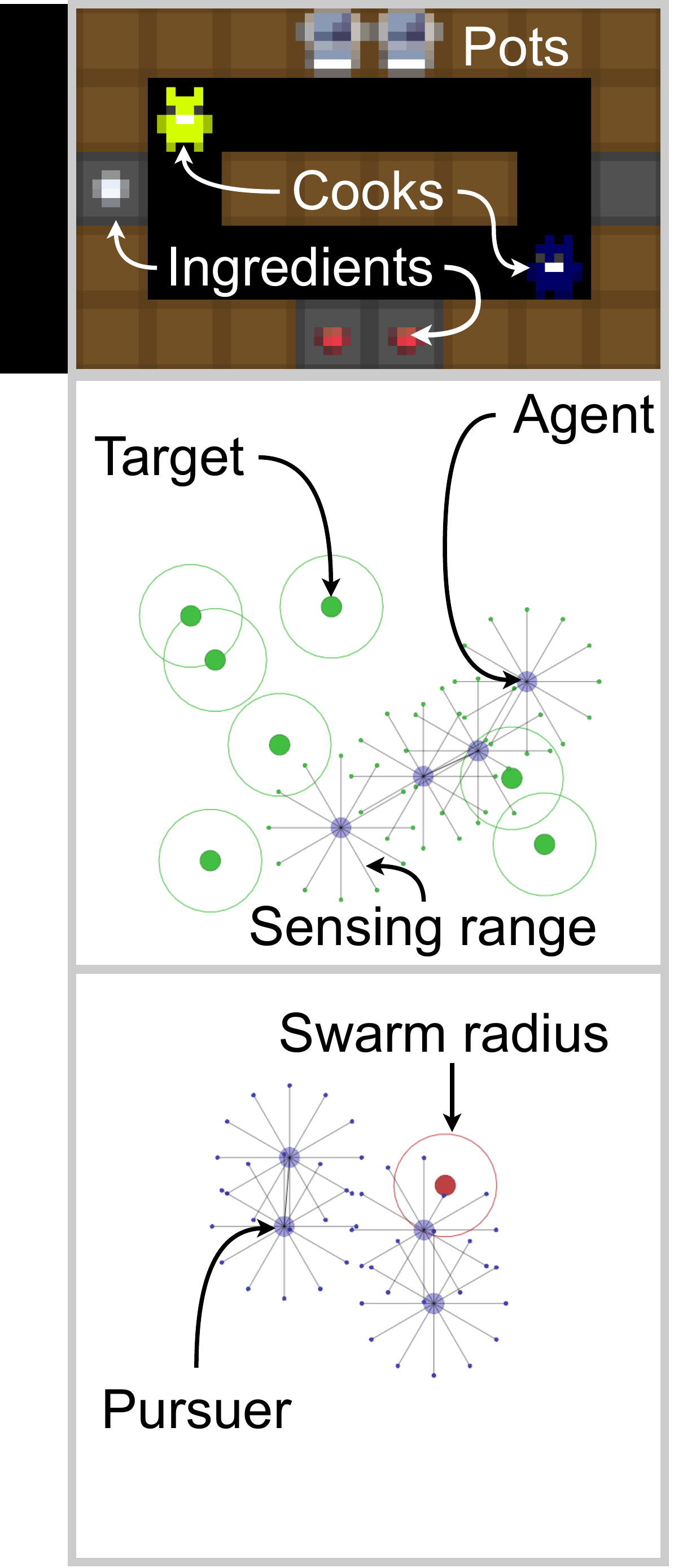}
    \caption{\textbf{Tasks.} Circuit (top), Discovery (middle), and Pursuit-Evasion (bottom).}
    \label{fig:tasks}
\end{wrapfigure}

We obtain our pre-training dataset by following a uniform random policy (Melting Pot) or uniform random sampling from the observation space (VMAS) for a million steps in the environment. With these samples, we train a task-agnostic communication strategy (Section \ref{sec:trainta}) and deploy it with our communication model (Section \ref{sec:commsmodel}). For all of our experiments, we optimise the policy for each agent using \textit{Proximal Policy Optimisation (PPO)} \citep{schulman_proximal_2017}. This is commonly referred to in multi-agent literature as \textit{Independent PPO (IPPO)}. However, we emphasise that our method is algorithm-agnostic. Any optimisation algorithm may be used with our method in place of IPPO.

\subsection{Performance on Novel Tasks}

In this experiment, we measure the converged return of our method (\textit{task-agnostic}) against two baselines as we learn policies for a variety of tasks. The \textit{task-specific} baseline simulates reusing communication strategies learned from other tasks. In a real use case, this is the only option to avoid training a new strategy if a task-agnostic strategy does not exist. For the task-specific baseline, we pre-train the set autoencoder using reinforcement learning while trying to learn a policy for a distinct but similar task in the same environment. We use an identical setup for pre-training the task-specific set autoencoder as when we evaluate the task-agnostic method. In contrast, the task-agnostic baseline uses random samples from the environment with reconstruction loss for pre-training. This approach lets us assess how well a task-agnostic method generalises compared to a task-specific one using the same architecture. The \textit{no-comms} baseline uses no communication strategy at all. For Melting Pot environments, we additionally utilise an image encoder, pre-trained in all tasks in an environment, which we use for every baseline. We evaluate our method, along with the baselines, on three distinct tasks (Figure \ref{fig:tasks}):

\textbf{Collaborative Cooking: Circuit.} In Melting Pot's collaborative cooking environment, the task is to complete recipes. For our task-agnostic strategy, we pre-train on the environment and deploy it to learn the \textit{Circuit} variant, where two agents must navigate around a circuit to access cooking pots and ingredients. The \textit{task-specific} variant uses a communication strategy which was learned in the \textit{Cramped} variant where agents must cook under tight space restrictions. 

\textbf{Discovery.} In this VMAS task, four agents must try to discover targets. To get a positive global reward, two out of four agents must position themselves within a small radius of a target to ``discover'' it. Together, agents must coordinate to discover targets as fast as possible. New targets continuously spawn as others are discovered. For this task, the \textit{task-specific} variant uses a communication strategy learned in VMAS' Flocking task where agents must learn to flock, much like birds do, around a moving target.

\textbf{Pursuit-Evasion.} The VMAS Pursuit-Evasion task is a find-and-intercept game. The agents are pursuers and they must catch an evading target. The visibility of the pursuers is limited. They must work together to find the target and collectively swarm around them in order to catch them as fast as possible. For this, the \textit{task-specific} variant uses a communication strategy learned in VMAS' Discovery task.

\begin{figure}
    \centering
    \includegraphics[width=0.8\linewidth]{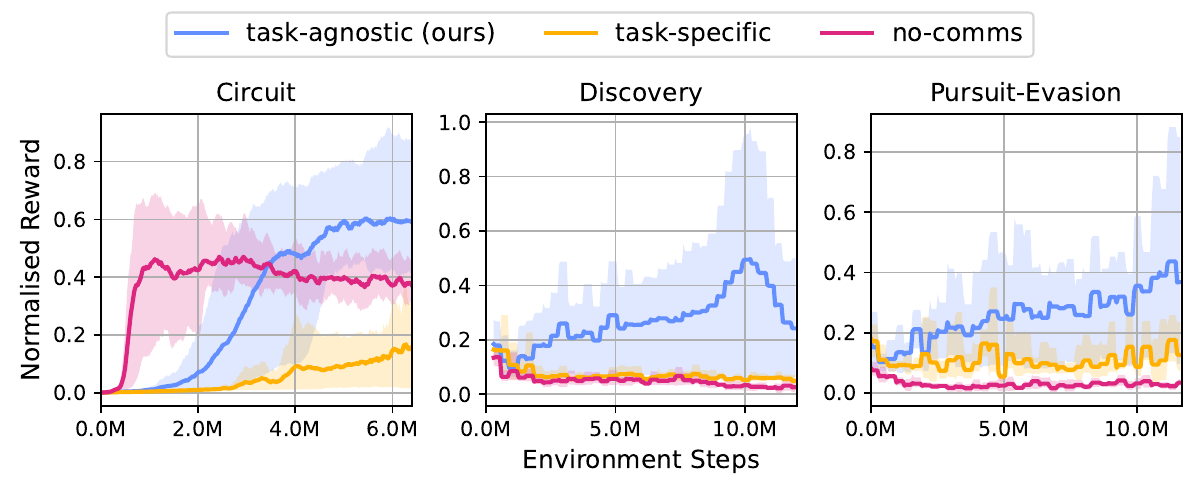}
    \caption{\textbf{Task-Agnostic communication strategies lead to greater rewards in novel tasks.} For each set of results, we report the mean and central 95\% interval over 5 seeds. We trained for 6.4 million environment steps in Melting Pot tasks, and 12 million environment steps in VMAS tasks. We used the same number of steps to pre-train the task-specific strategy on a similar task.}
    \label{fig:agnostic}
\end{figure}

We see a significant improvement in return when using task-agnostic strategies (Figure \ref{fig:agnostic}). In \textbf{Circuit}, the task-specific baseline fails to achieve a mean reward much higher than the starting reward. The no-comms baseline stops improving after around 1M steps. In contrast, our task-agnostic method continuously improves, outperforming both baselines.

In \textbf{Discovery}, both the task-specific and no-comms baselines fail to learn a useful representation. Meanwhile, the task-agnostic strategy produces a better policy almost immediately, gradually improving and peaking after around 10M steps. We outperform the two baselines from just after the start and through to the end of training.

In \textbf{Pursuit-Evasion}, while the task-specific baseline appears to outperform the no-comms baseline, much like Discovery, both plateau after a small number of training steps. The task-agnostic outperforms both baselines.

The results show that task-agnostic communication strategies consistently enable agents to leverage communication without relearning the communication strategy. This is useful in the real world, where cooperative robots engage in a variety of tasks in a shared environment.

\subsection{Generalisation with Out-Of-Distribution Numbers of Agents}
\label{sec:ood-agents}

In real-world situations, new agents may join the multi-agent team during execution to support other agents. To cope with such dynamic scenarios, we investigate how pre-trained reconstructions of the global state generalise to an out-of-distribution (OOD) number of agents. Since the PISA encoder's latent state is fixed-size, input cardinality is independent of output dimensionality---we can encode a set of any cardinality into a constant size latent vector. Therefore, in this experiment, we measure the performance of our communication strategy when we have more agents, and hence larger sets, than seen during pre-training.

\begin{figure}[t]
    \centering
    \includegraphics[width=0.8\linewidth]{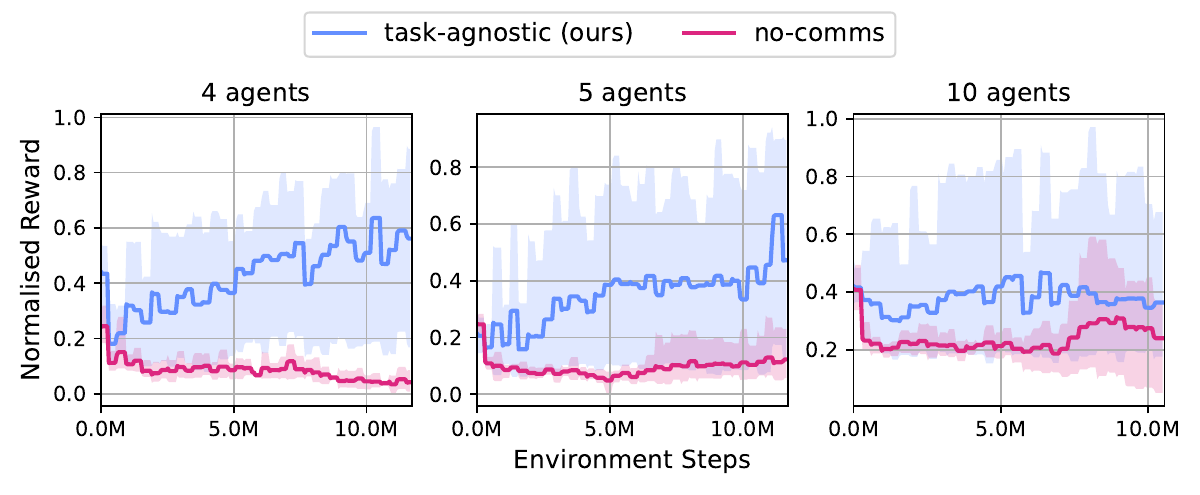}
    \caption{\textbf{Our task-agnostic strategy scales out-of-distribution.} We pre-trained the communication strategy with 1, 2, and 3 agents for 1M environment steps and trained the policy for 12M environment steps. For each set of results, we report mean and central 95\% interval over 5 seeds.}
    \label{fig:experiment2}
\end{figure}

We pre-train our set autoencoder in the Discovery environment with 1, 2, and 3 agents. Then, we train and evaluate a policy on more agents than seen during pre-training. Under these conditions, in Figure \ref{fig:experiment2}, we show that our method still significantly outperforms the baseline when we learn a policy with 4 and 5 agents, going beyond the number of agents that we pre-trained our communication strategy with. This is evidence that our approach can elegantly handle changes in connectivity (e.g. from communication disruptions) and can support variable numbers of agents without fine-tuning. Even once we reach 10 agents, although the performance gap is smaller, we continue to outperform the no-communication baseline. At this point, we are far outside the training distribution.

\subsection{Detecting Out-Of-Distribution Events}

We often want to detect out-of-distribution events in an environment. For example, if a disturbance occurs (e.g., humans entering a robot-only operating area, or adversarial agents accessing the communications network), we want agents to safely halt or take appropriate actions. In this experiment, we detect OOD occurrences by comparing the reconstruction loss during training and at runtime.

\begin{figure}[t]
\begin{subfigure}{0.45\linewidth}
    \centering
    \includegraphics[width=\linewidth]{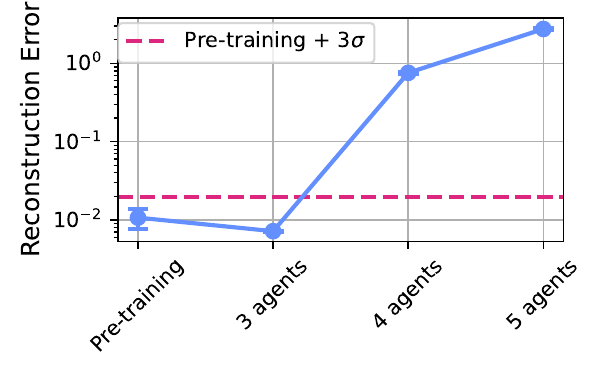}
    \caption{\textbf{Detecting OOD agents.} We report the mean set reconstruction error of PISA over the last 10 iterations of policy training.}
    \label{fig:experiment3}
\end{subfigure} \hfill
\begin{subfigure}{0.45\linewidth}
    
    \includegraphics[width=\linewidth]{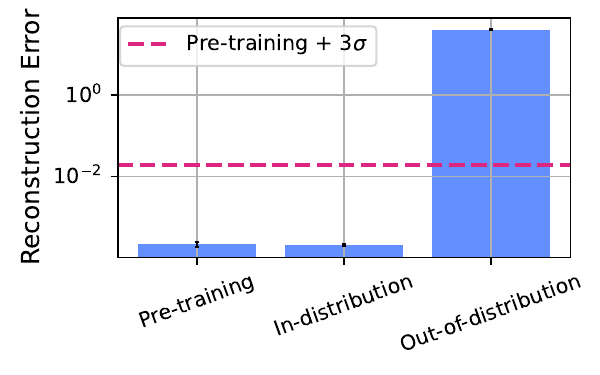}
    \caption{\textbf{Detecting OOD observations.} We measure the mean set reconstruction error of PISA over the last 10 iterations of policy training.}
    \label{fig:experiment32}
\end{subfigure}
\caption{\textbf{Detecting out-of-distribution states.}}
\end{figure}

In Figure \ref{fig:experiment3}, we show how the set reconstruction error changes when we deploy a communication strategy pre-trained with 1, 2, and 3 agents on Discovery (as in Section \ref{sec:ood-agents}) to train policies with 3 (in-distribution), 4 (OOD), and 5 (OOD) agents in the same environment. A threshold set by the pre-training loss mean plus three standard deviations easily detects the OOD agent counts.

Similarly, we can detect OOD observations. In the Collaborative Cooking environment, when we fix one of the agents to receive only Gaussian noise observations (OOD), the loss exceeds our threshold (Figure \ref{fig:experiment32}).

\subsection{Limitations.} For simplicity, we use full connectivity between agents. However, this is not a technical limitation since it can be overcome by propagating information via aggregation (e.g. aggregating sets of PISA encodings with another PISA). Additionally, collecting pre-training samples with a scheme such as curiosity-driven exploration \citep{pathak_curiosity-driven_2017} could lead to more efficient representations of the Markovian state from the autoencoder as it samples sparse states more frequently.

\section{Related Work}
Prior MARL papers have neglected the inefficiency of relearning communication strategies for distinct tasks. We addressed this by introducing task-agnostic communication strategies that can be shared for all tasks within an environment. %

\textbf{Differentiable Communication.} Differentiable models of communication optimise messages with respect to the objective on-policy. Some typical examples are DIAL \citep{foerster_learning_2016}, CommNet \citep{sukhbaatar_learning_2016}, TarMAC \citep{das_tarmac_2019}, \citet{eccles_biases_2019}, HetGPPO \citep{bettini_heterogeneous_2023}, EPC \citep{long_evolutionary_2020}, SPC \citep{wang_towards_2023}, and \citet{abdel-aziz_cooperative_2023}. All of these methods employ task-specific communication strategies and thus require optimising the strategy for each distinct task. In contrast, our method is task-agnostic---tasks within an environment can share our pre-trained communication strategies. While some of these works support variable numbers of agents (population-invariant communication), DIAL, TarMAC, \citet{eccles_biases_2019} and \citet{abdel-aziz_cooperative_2023} do not. Our approach supports population-invariant communication \textit{in addition to} task-agnostic communication through a fixed-size latent state in the autoencoder.

\textbf{Self-Supervised Communication.} Several recent works have used self-supervised and contrastive objectives to train differentiable communication strategies \citep{lin_learning_2021, guan_efficient_2022, lo_learning_2022, lo_learning_2023}. All of these methods learn the communications policy \emph{online}, biasing the communications towards a specific objective, while we learn it offline without any bias. Hence, they are not task-agnostic strategies. Furthermore, none of these methods support variable numbers of agents as ours does.

\textbf{Pre-training in RL.} Contemporary works in RL have utilised pre-training to leverage prior knowledge when training policies \citep{cruz_pre-training_2017, singh_parrot_2021, yang_representation_2021, schwarzer_pretraining_2021, seo_reinforcement_2022}. Fundamentally, they all attempt to learn representations that are useful for solving the underlying MLP through various unsupervised methods. While all of these works focus on single-agent RL, we utilise pre-training to improve the efficiency of MARL.

\section{Conclusion}
We proposed task-agnostic communication strategies to eliminate the inefficiency of task-specific multi-robot communication. By using a set autoencoder to reconstruct the global state from local observations, our approach is guaranteed to converge under modest assumptions, with an upper bound on regret due to approximating the Markovian state. Empirically, it outperforms task-specific strategies in novel tasks, scales to more agents than in pre-training, and detects out-of-distribution events during policy training using pre-training losses.

As it stands, our method is adaptable to various learning paradigms, not just RL, because it pre-trains communication strategies in a self-supervised manner. As we avoid end-to-end training, we also expedite RL policy training by tuning fewer weights. Additionally, having pre-trained an autoencoder, our policy can use sparse reward signals more efficiently as it does not need to learn environment-specific features. Lastly, our method opens up new applications for key real-world robotics tasks such as allowing changing policies at runtime, or running heterogeneous policies on different collaborative robots.

\section{Acknowledgements}
This work is supported by ARL DCIST CRA W911NF-17-2-0181 and European Research Council (ERC) Project 949940 (gAIa). We gratefully acknowledge their support. We thank Ryan Kortvelesy and Matteo Bettini for their assistance, as well as Edan Toledo, Yonatan Gideoni, and Jonas Jürß for reviewing early drafts.

\bibliographystyle{spbasic}
\bibliography{references, newrefs}

\appendix
\clearpage
\section{Modifications To Tasks And Environments}
\label{app:mods}
As alluded to in Section \ref{sec:setup}, we enforce that Melting Pot tasks are Dec-MDPs with a global reward $R$. In the default case, an agent $i$ receives an individual reward $r_i$. We modify this such that each individual agent reward $r_i$ is summed to make a global reward $R = \sum_i^n r_i$ and this global reward is shared between all agents instead of them receiving their individual reward.

In all collaborative cooking task variants, we enable a pseudo-reward that rewards agents with $+1$ for picking up a tomato and putting it in a pot. Without this pseudo-reward, these tasks have extremely sparse rewards that make them highly challenging to learn. This shaped reward assists in learning on these typically very difficult tasks.

In VMAS, we modify the observation spaces of Discovery and Flocking to be identical by replacing Discovery's redundant additional position vector feature with a two-dimensional zero vector and reducing the number of LiDAR rays from 15 to 12. This allows us to use communication strategies learned in Flocking on Discovery. We argue that Flocking and Discovery are two tasks in the same environment because the obstacles in Flocking can be treated as targets to derive the Discovery task from Flocking (along with a change of reward).

Furthermore, we designed and created Pursuit-Evasion ourselves based on both Flocking and Discovery. This is another task that we argue is in the same environment. It can be derived from Flocking if we treat Flocking's target as a robber, and the agents as the police after modifying the reward. By extension, it can also be derived from Discovery.

\change{\section{Permutation-Invariant Set Autoencoder Architecture}

The set autoencoder architecture is based on \citet[Figure 1]{kortvelesy_permutation-invariant_2023}. It encodes a variable-sized set of elements $\{x_1, x_2, \dots, x_n\}$ where $x_i \in \mathbb{R}^n$ into a fixed-size permutation-invariant latent state $z \in \mathbb{R}^z$. It is trained with a self-supervised reconstruction objective to decode the latent state and recover the set.

\textbf{Encoder.} The encoder takes the input set $\{x_1, x_2, \dots, x_n\}$ and maps each element to a key $k_i$ according to some criterion and encodes the keys using a network $\psi_{\mathrm{key}}$. Simultaneously, the encoder takes the input set elements and also encodes them into values using a separate network $\psi_{\mathrm{val}}$. The encoder then takes the element-wise product of the corresponding key and value embeddings and sums them all. Finally, a cardinality embedding $\lambda_{\mathrm{enc}}(n)$ is added to this sum to form the final latent state $z$.

\textbf{Decoder.} The decoder takes the latent state $z$ and predicts the cardinality of the set with a network $\lambda_{\mathrm{dec}}$. The predicted cardinality is used to create a set of keys as in the encoder and the keys are mapped to queries by a network $\phi_{\mathrm{key}}$. Each query is element-wise multiplied by the latent state and a final decoder network $\phi_{\mathrm{dec}}$ recovers the set from these embeddings.

Further details may be found in \citet{kortvelesy_permutation-invariant_2023}. The hyperparameters used in our work are detailed in Appendix \ref{app:pretrainhp}.
}

\section{Policy Network Architectures}
\label{app:arch}
When training to solve Melting Pot tasks, we independently train the policy of each agent with PPO. For each agent, we have independent three-layer MLPs as our policy and value networks. The policy network's hidden layer is 128 neurons wide, while the value network's hidden layer is 1024 neurons wide. We initialise the last layer of the policy network and value network using a normal distribution with zero mean and 0.01 standard deviation in line with the suggestions made by \citet{andrychowicz_what_2020}.

Unlike Melting Pot, as no heterogeneous behaviour is required for our VMAS tasks, we train a policy that's shared between all agents with PPO. For each agent, we have independent three-layer MLPs as our policy and value networks. For Discovery, the policy and value networks have a 256-wide hidden layer while for Pursuit-Evasion, the hidden layers are 512-wide. We initialise the last layer of the policy and value networks with the same normal distribution as we use for Melting Pot.

\section{Training Hyperparameters}
\label{app:trainhp}
Our training hyperparameters are dependent on the multi-agent suite and task and are described in Table \ref{tab:mphp} and \ref{tab:vmashp}. We always use fixed seeds 0-4 for every experiment. Specifically for Pursuit-Evasion, we use the defaults for the parameters in Table \ref{tab:vmashp} except for train batch size, SGD minibatch size, training iterations, and rollout fragment length.

\vspace{1em}
\begin{table}[h]
    \centering
    \begin{tabular}{lc}
        \toprule
        \textbf{Parameter} & \textbf{Value} \\
        \midrule
        Train batch size & 6400 \\
        SGD minibatch size & 128 \\
        Training iterations & 1000 \\
        Rollout fragment length & 100 \\
         \bottomrule
    \end{tabular}
    \caption{Melting Pot training hyperparameters.}
    \label{tab:mphp}
\end{table}
\vspace{1em}

\begin{table}[h]
    \centering
    \begin{tabular}{lc}
        \toprule
        \textbf{Parameter} & \textbf{Value} \\
        \midrule
        Train batch size & 60000 \\
        SGD minibatch size & 4096 \\
        Training iterations & 200 \\
        Rollout fragment length & 125 \\
        KL coefficient & 0.01 \\
        KL target & 0.01 \\
        $\lambda$ & 0.9 \\
        Clip & 0.2 \\
        Value function loss coefficient & 1 \\
        Value function clip & $\infty$ \\
        Entropy coefficient & 0 \\
        $\eta$ & 5e-5 \\
        $\gamma$ & 0.99 \\
         \bottomrule
    \end{tabular}
    \caption{VMAS training hyperparameters.}
    \label{tab:vmashp}
\end{table}

\section{Pre-Training Hyperparameters}
\label{app:pretrainhp}
For Melting Pot environments we train an image encoder in addition to PISA. Observations are first encoded with the image encoder before this embedding is passed to PISA. For the image encoder, we use a 3-layer CNN encoder and decoder as specified in Table \ref{tab:mpimageencoder}. Our training data is gathered from observations of all agents generated with a uniform random policy rolled out over 1M environment steps. We train the image encoder with a mini-batch size of 32 for approximately 1000 iterations or until the loss has clearly converged.

\vspace{1em}
\begin{table}[h]
    \centering
    \begin{tabular}{llcccccc}
        \toprule
        \textbf{Layer} & \textbf{Type} & \textbf{In ch.} & \textbf{Out ch.} & \textbf{Kernel} & \textbf{Stride} & \textbf{Padding} & \textbf{Activation} \\
        \midrule
        Encoder & Conv2D & 3 & 16 & 3 & 2 & 1 & ReLU \\
        & Conv2D & 16 & 32 & 3 & 2 & 1 & ReLU \\
        & Conv2D & 32 & 64 & 3 & 2 & 1 & ReLU  \\
        & Linear & 1600 & 128 & - & - & - & - \\
        \midrule
        Decoder & Linear & 128 & 1600 & - & - & - & -\\
        & ConvTranspose2D & 64 & 32 & 3 & 2 & 1 & ReLU \\
        & ConvTranspose2D & 32 & 16 & 3 & 2 & 1 & ReLU \\
        & ConvTranspose2D & 16 & 3 & 3 & 2 & 1 & Sigmoid \\
        \bottomrule
        
    \end{tabular}
    \caption{Melting Pot image autoencoder architecture.}
    \label{tab:mpimageencoder}
\end{table}
\vspace{1em}

For the set autoencoder, we use the default implementation of PISA provided in the author's repository\footnote{\url{https://github.com/Acciorocketships/SetAutoEncoder/tree/main}}. We train PISA with a latent dimension of 256 with a batch size of 32 for 15000 iterations or until the loss has clearly converged.

Unlike Melting Pot, we do not train an image encoder for VMAS environments as observations are already feature vectors. We train PISA with a latent dimension of 72 with a batch size of 256 for 15000 iterations where the loss has clearly converged.

Since VMAS environments are extremely simple, we find that uniformly randomly sampling from the observation space to generate pre-training data works well. This leads to learning a strong autoencoder where the reconstruction loss is very small during policy training. Hence, we use this method in our final results rather than a uniform random policy. Both lead to task-agnostic communication strategies as they are reward-free.

\section{Choosing MARL Benchmarks}
\label{app:benchmarks}
\change{While other well-known MARL benchmarks exist, we choose not to use these as they either do not require communication to solve \citep{samvelyan_starcraft_2019}, lack sufficient task variation \citep{samvelyan_starcraft_2019, ellis_smacv2_2022, kurach_google_2020}, or are not Dec-MDP/POMDPs \citep{kurach_google_2020}.

While SMAC \citep{samvelyan_starcraft_2019} is commonly used in prior literature, many of its environments can be solved with open-loop policies (i.e. with observations of just the agent ID and time step) \citep{ellis_smacv2_2022}. As a result, communication is not necessary to solve it. While SMACv2 \citep{ellis_smacv2_2022} resolves some of these issues, the objective remains simply to kill all the enemy agents. Consequently, this environment does not have enough task variation to test task-agnostic communication, the main contribution of this paper. GRF \citep{kurach_google_2020} has similar issues.

Melting Pot \citep{agapiou_melting_2023} represents similarly challenging tasks. Like SMAC and GRF, it features high-dimensional pixel-based observations and complex objectives. It is a new, but state-of-the-art benchmark. VMAS \citep{bettini_vmas_2022} is also a suitably challenging benchmark. The visualisations in VMAS appear simple, but the dynamics are complex, going beyond kinematics by simulating elastic collisions, rotations, and joints. Thus, while the environments are conceptually basic, VMAS represents a \textit{realistic} challenge to agents.

}

\section{Computation, Hardware, and Implementation Details}
\label{app:compute}
We implemented our work with the Ray RLLib library (version 2.1.0 for VMAS and 2.3.1 for Melting Pot) and wrote all our models with the PyTorch framework. Our models and policies were primarily trained on individual NVIDIA A100 GPUs with 40GiB of memory and NVIDIA RTX 2080Ti GPUs with 11GiB of memory. Experiments were conducted with 5 workers for VMAS with 32 vectorised environments and 2 workers for Melting Pot. In each case, we used a single driver GPU while environment simulations were carried out on CPU. Training a policy for 12M environment steps on a VMAS task took approximately 6-12 hours, while 6.4M environment steps on Melting Pot took about 18-24 hours. Pre-training the image encoder took about 6 hours and pre-training PISA took about 1 hour for VMAS and 6 hours for Melting Pot.

\end{document}